\documentclass{aa}  
\usepackage{natbib}
\usepackage{graphicx}
\usepackage{txfonts}

\begin{document}

\title{The Binding Energies of Atoms on Amorphous Silicate Dust: A Computational Study
}

\author{Kristoffer Hansson\inst{1}
    W. M. C. Sameera,*\inst{1}\textsuperscript{,}\inst{2}
    Clarke J. Esmerian,*\inst{2} 
    Duncan Bossion,\inst{3} 
     Stefan Andersson,\inst{1}\textsuperscript{,}\inst{4}
     Susanne Aalto,\inst{2}
      Wouter Vlemmings,\inst{2}
    Kirsten K. Knudsen,\inst{2}
    \and
    Gunnar Nyman*\inst{1}
}

\institute{\textit{Department of Chemistry and Molecular Biology, University of Gothenburg, SE-412 96 Gothenburg, Sweden}\\
    \email{nyman@chem.gu.se}
    \and
    \textit{Department of Space, Earth and Environment, Chalmers University of Technology, SE-412 96 Gothenburg, Sweden}\\
    \email{chamil@chalmers.se, clarke.esmerian@chalmers.se}
    \and
    \textit{Institute of Physics of Rennes, UMR-CNRS 6251, University of Rennes, 35000 Rennes, France}\
    \and
    \textit{SINTEF, P.O. Box 4760 Torgarden, NO-7465 Trondheim, Norway}
}

\date{Received ??, 2025; accepted ??, 2025}

\abstract
   {We investigate the binding energies of atoms to interstellar dust particles, which play a key role in their growth and evolution, as well as for the chemical reactions on their surfaces.}
   {We aim to compute the binding energies of abundant atoms in the interstellar medium (C, N, O, Mg, Al, Si, S, Ca, Fe, and Ni) to silicate dust.}
  % methods heading (mandatory
   {We used the Geometries, Frequencies, and Non-covalent Interactions Tight Binding (GFN1-xTB) method to compute the binding energies. An FeMgSiO\textsubscript{4} periodic surface model, containing 81 local minima on the surface, was used.}
   {A range of binding energies was found for each element. The median of the binding energies follows the order Si (14.8 eV) > Al (12.8 eV) > Ca (12.7 eV) > C (9.5 eV) > O (8.1 eV) > N (6.2 eV) > Fe (6.0 eV) > S (5.2 eV) > Mg (2.4 eV). The probability distribution of binding energies for each element except Ca is statistically consistent with a log-normal distribution. }
   {In general, Si, Ca, and Al atoms have large binding energies. Thus, these atoms can stay on the silicate dust particles at high temperatures. The binding energies of the other atoms, C, N, O, Mg, S, Fe and Ni, are relatively weak. However, the computed binding energies for these elements are still far stronger than the energies associated with dust temperatures typical of the ambient interstellar medium (ISM), suggesting that silicate grains are generally stable against sublimation. We estimate sublimation temperatures for silicate grains to range from 1600 K to 3000K depending on assumed grain size and lifetime. These binding energies on silicate dust grains, estimated from first principles for the first time, provide invaluable input to models of dust evolution and dust-catalyzed chemical reactions in the interstellar medium.}
\keywords{binding energies of atoms -- silicate dust --interstellar medium -- GFN1-xTB --}
\authorrunning{Short Author Name} \titlerunning{Short Title}
\maketitle
%________________________________________________________________
\section{Introduction}

   Interstellar dust plays a crucial role in the formation of molecular clouds, where stars and planets are formed. Thus, the composition of dust particles, their origin, and their evolution are fundamental questions in astrophysics \citep{Tielens_2005,40fe489fc8f9431aaf81f917abb41797,10.3389/fspas.2022.908217}. However, fundamental uncertainties remain in all of these areas.
   
   In the 1940s, van de Hulst  proposed making particles out of atoms (e.g. H, O, C, and N) in space, where the atoms combined on a surface of frozen saturated molecules, the so-called “dirty ice” model \citep{van1943vorming,oort1946gas}. An alternative dust material consisting of small graphite flakes formed around cool carbon stars was proposed by \citet{10.1093/mnras/124.5.417}. Soon after, the same authors proposed silicate and graphite grains (with and without ice mantles) as possible dust materials \citep{10.1093/mnras/126.1.99}. Hoyle \& Wickramasinghe further indicated that the graphite along with silicate grains are major interstellar dust species, which can be formed around oxygen-rich giant stars \citep{hoyle1969interstellar}. Later, uncoated graphite, enstatite, olivine, silicon carbide, iron, and magnetite were proposed as interstellar dust analogues \citep{mathis1977size,draine1984optical}. Also, polycyclic aromatic hydrocarbon (PAH) molecules were proposed as interstellar dust constituents \citep{desert1990interstellar,siebenmorgen1992dust,dwek1997detection,draine2001infrared,li2001infrared,li2002infrared,siebenmorgen2014dust}. 
   
   In modern models of cosmic dust, different grain populations, such as graphite, olivine, silicate, PAH, and amorphous carbons are possible \citep{compiegne2011global,draine1984optical,siebenmorgen1992dust,dwek1997detection,draine2001infrared,draine2007infrared,li2001infrared,li2002infrared,greenberg1983far,siebenmorgen2014dust}. Some authors argue that interstellar dust consists rather of silicate grains with hydrogenated amorphous carbon mantles \citep[e.g.][]{greenberg1986role,duley1987interstellar,jones1987interstellar,jones1990structure,jones2013evolution,jones2014cycling,duley1989hydrogenated,jones1990carbon,sorrell1991annealed,li1997unified,zonca2011modelling,cecchi2014observational,kohler2014hidden}. These core-mantle (CM) interstellar grains are now a widely used concept, where CM accounts for dust variations (i.e., carbonaceous mantle composition and/or its depth) through environmentally driven changes \citep{jones1987interstellar,jones1990structure,jones2013evolution,jones2016mantle,kohler2014hidden,duley1989hydrogenated,zonca2011modelling,cecchi2014observational,cecchi2014extragalactic,ysard2015dust,ysard2016mantle}.
   
   Atoms in the gas phase can adsorb on dust and contribute to the growth of the dust particle if the binding energy of the atoms on the dust particles is strong. It is thought that this process is a central, if not dominant, source of the interstellar dust content in galaxies over cosmic history \citep[e.g.][]{Esmerian2022, Esmerian2024}. If the binding energy is too weak, the atoms may diffuse on the dust particle or they may desorb. In addition, the temperature at which dust grains sublimate due to the thermal expulsion of atoms from their surfaces is also determined by these binding energies. Indeed, observations indicate that dust also survives in exceptionally harsh environments, most notably in the circumnuclear tori of active galactic nuclei (AGN) where grains experience temperatures of ~1000–2000 K \citep[e.g.][]{Nenkova2008,Netzer2015}. Therefore, this quantity is essential for understanding the growth and evolution of dust particles in the interstellar medium (ISM). In this study, we compute the binding energies of atoms on a silicate interstellar dust particle. 
   
   The precise composition of interstellar silicate material is uncertain because it must be determined from a combination of only a few spectral features, assumptions about the optical properties of candidate materials (which are uncertain), and estimates of total elemental abundances and their depletion from the gas phase (which are also uncertain). \citet{min2007shape} proposed that the overall silicate composition is Mg\textsubscript{1.32}Fe\textsubscript{0.1}SiO\textsubscript{3.45}, while \citet{Poteet_2015} favor the composition of Mg\textsubscript{1.48}Fe\textsubscript{0.32}SiO\textsubscript{3.79}. \citet{fogerty2016silicate}  proposed the overall composition of Mg\textsubscript{1.37}Fe\textsubscript{0.18}Ca\textsubscript{0.002}SiO\textsubscript{3.55}. A recent work by \citet{draine2021dielectric} proposed the overall composition as Mg\textsubscript{1.3}(Fe,Ni)\textsubscript{0.3}SiO\textsubscript{3.6}. A lack of characteristic spectral features suggests that these silicate grains are not crystalline but rather amorphous in their molecular geometries \citep{do2020crystalline, Kemper_2005, demyk1999chemical}.
   
   As the composition of dust remains uncertain, we employed the known crystal structure of the silicate material FeMgSiO\textsubscript{4} as a model system to create a periodic slab structure of a dust particle, which we amorphized by heating to 5000 K in a molecular dynamics simulation. We then computed the binding energies of C, N, O, Mg, Si, S, Al, Ca, Fe, and Ni atoms, which have relatively high gas-phase abundances and are therefore the most likely contributors to interstellar dust growth. To the best of our knowledge, this is the first time these binding energies are estimated from first-principles calculations. A range of binding energies is found for each element, providing invaluable data for the predictions of interstellar dust properties and processes in the ISM. We describe our computational methods in Section \ref{sec:methods}, our results in Section \ref{sec:results}. We discuss their implications in Section \ref{sec:discussion} and conclude in Section \ref{sec:conclusion}.
   
%__________________________________________________________________

\section{Computational Methods}\label{sec:methods}

    \begin{figure}
   \centering
   \includegraphics[width=\hsize]{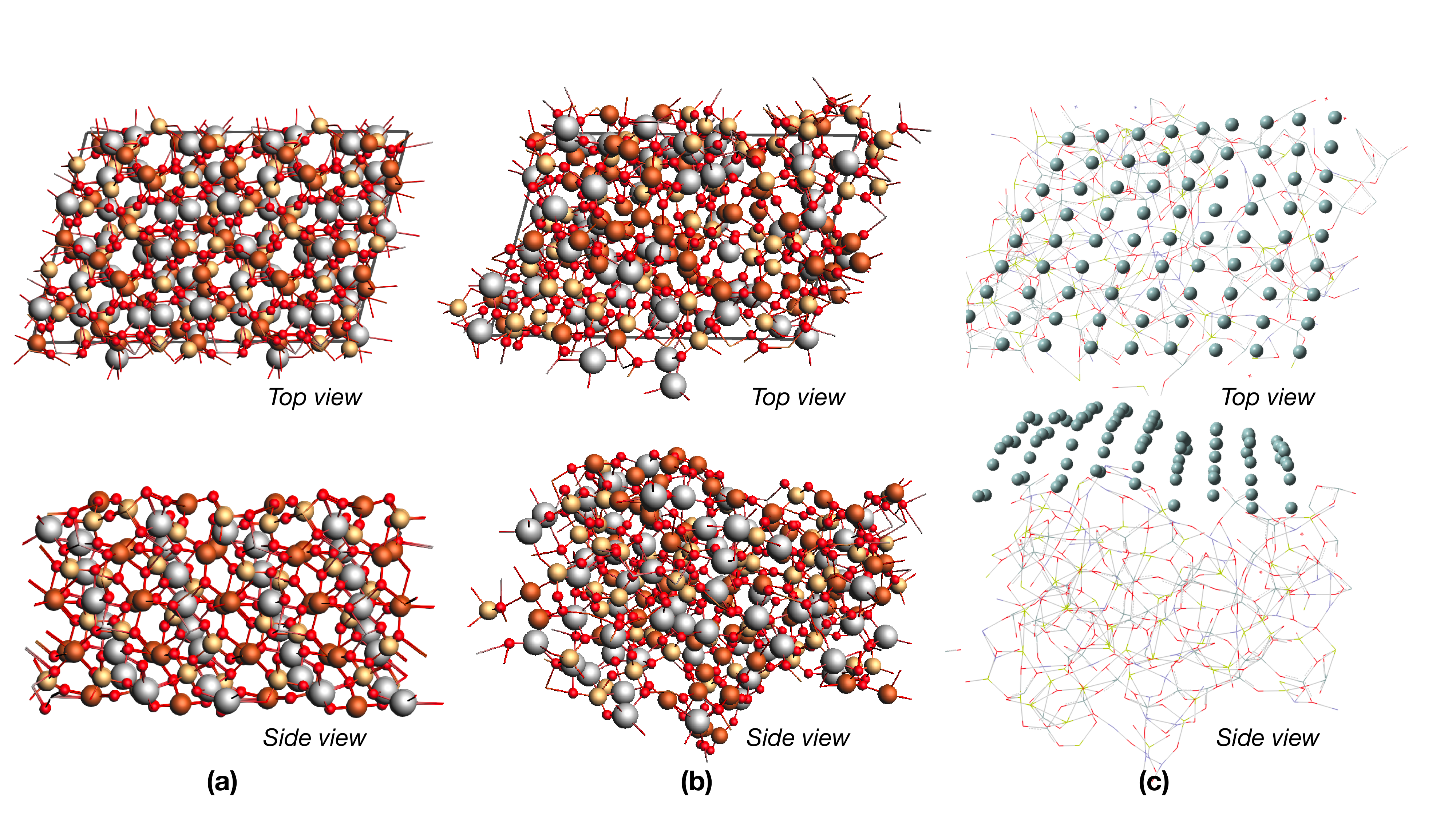}
      \caption{{\bf Column (a):} Unit cell of the optimized FeMgSiO\textsubscript{4} periodic slab before the MD simulation of heating to 5000K to amorphize the structure. {\bf Column (b):} FeMgSiO\textsubscript{4} periodic slab structure after amorphization proceedure. {\bf Column (c):} 81 grid points on which we calculate binding energies.}
         \label{FigVibStab}
   \end{figure}

The calculations were performed using the GFN1-xTB (Geometries, Frequencies, and Non-covalent Interactions Tight Binding) method \citep{grimme2017robust}, as implemented in the AMS DFTB 2024.105 SCM program \citep{te2001chemistry}. Starting from the crystal structure of FeMgSiO\textsubscript{4} \citep{chatterjee2009site}, a periodic slab structure was created by slicing the bulk crystal along the plane (111) in the crystal lattice, where the term (111) corresponds to a specific crystallographic plane that intersects the x, y, and z axes of a cubic crystal at equal distances. The unit cell, the smallest repeating unit in a crystal lattice, has an a-axis length of 15.61 Å, a b-axis length of 23.15 Å, and an angle of 104.9$^{\circ}$ between the a-axis and b-axis. The unit cell consist of 64 Fe\textsuperscript{2+} ions, 64 Mg\textsuperscript{2+} ions, 64 Si\textsuperscript{4+} ions and 256 O\textsuperscript{2-} ions (Figure 1a). This structure therefore represents a small surface section of typical silicate cosmic dust grains, of which the vast majority are $\gg 10~\AA$ and the most numerous are of order 0.1$\mu{\rm m} = 1000~\AA$ \citep{HensleyDraine2023}. This justifies the plane-parallel geometry of our computational set-up. 

The periodic slab structure was relaxed to get an energetically favorable configuration using the GNF1-xTB method. The resulting structure was used as the initial configuration for a molecular dynamics (MD) simulation at 5000 K. During the MD simulation, the atomic positions of the periodic slab structure evolved according to Newton's equations of motion, enabling the time-dependent structural fluctuations to converge to an equilibrium configuration. A relatively high temperature (i.e., 5000 K) was set to induce atomic mobility that promoted the formation of an amorphous structure. 

In the MD simulation, temperature control is essential. For this purpose, the Berendsen thermostat \citep{berendsen1984molecular} was employed, which can maintain the temperature by weakly coupling the simulated system to an external heat bath. Moreover, the thermostat works by scaling the velocities of atoms in the system, allowing the temperature to gradually approach the target temperature. For the MD simulation, a timestep of 0.25 fs was used to achieve stable and accurate integration of atomic trajectories. As the simulation progresses, the total energy gradually stabilizes around 700 fs, where the temperature oscillates around 5000 K (Fig A.1). 

After the MD simulation, we performed a geometry optimization to find a stable configuration of the resulting structure, which was used as the FeMgSiO\textsubscript{4} dust particle model (Figure 1b). After that, 81 uniformly spaced grid points were selected approximately 4 Å above the optimized dust particle model surface (Figure 1c). We chose that distance as it is longer than the typical interatomic spacing, thus, weak interactions between the atom sitting on the grid and the dust particle guide the atom to find an energy minimum during the geometry relaxation. The binding energy of the atom was computed using the following formula:
\begin{equation}
    E_{\rm binding} = |E_{\rm dust-atom}| - |E_{\rm dust}| - |E_{\rm atom}|
\end{equation}

In this formula, \textit{E}\textsubscript{dust-atom} is the total potential energy of the optimized dust particle with the atom attached to it. The \textit{E}\textsubscript{dust} term represents the potential energy of the optimized dust particle model, and \textit{E}\textsubscript{atom} is the potential energy of the atom.  
%__________________________________________________________________

\section{Results} \label{sec:results}

\subsection{Binding Energy Distributions}

 The summary statistics of the binding energy distribution for each element are given in Table 1. The geometries of the optimized local minima of the selected structures are shown in Figure 1. The computed mean binding energies follow the order Si (15.31 eV)  > Ca (13.52 eV) > Al (12.79 eV) > C (9.20 eV) > O (8.10 eV) > N (6.37 eV) > Fe (5.92 eV) > S (5.17 eV) > Mg (2.57 eV). Si is typically the most strongly bound atom, while Mg is typically the weakest, although there is substantial spread in the values for any one element. 

The computed data showed a range of bonding energies for Si on the FeMgSiO\textsubscript{4} dust particle model, where 10.40 eV is the weakest binding energy, and 22.21 eV is the strongest binding energy. The geometries of the local minima suggest that in almost all cases the Si atom does not stay on the dust particle surface. Instead, the Si atom forms several chemical bonds with the atoms on the surfaces, moves into the particle, and remains near the surface. The binding energy is the lowest when the Si atom establishes two bonds with the surface O atoms (Figure 2a). We have selected the binding site that gives the median binding energy, where the Si atom forms four bonds with the dust particle surface atoms (three O and one Fe, Figure 2b). The computed binding energy is the strongest when Si forms four bonds with the surface O atoms (Figure 2c). 

In the case of the Ca atom, four to nine chemical bonds can be formed with O atoms on the surface of the dust particle (Figure 2d-f). Thus, the computed binding energies are strong (the weakest binding energy is 10.28 eV, and the strongest binding energy is 22.28 eV). As a result, the Ca atom can also move into the dust particle (while also remaining near the surface). The Al atom also showed a relatively strong binding energy, where it can create three chemical bonds with the O atoms of the dust particle (Figure 2g-i). Thus, the Al atom sits either on the surface or goes into the dust particle and stays near the surface. 

   \begin{figure}
   \centering
   \includegraphics[width=\hsize]{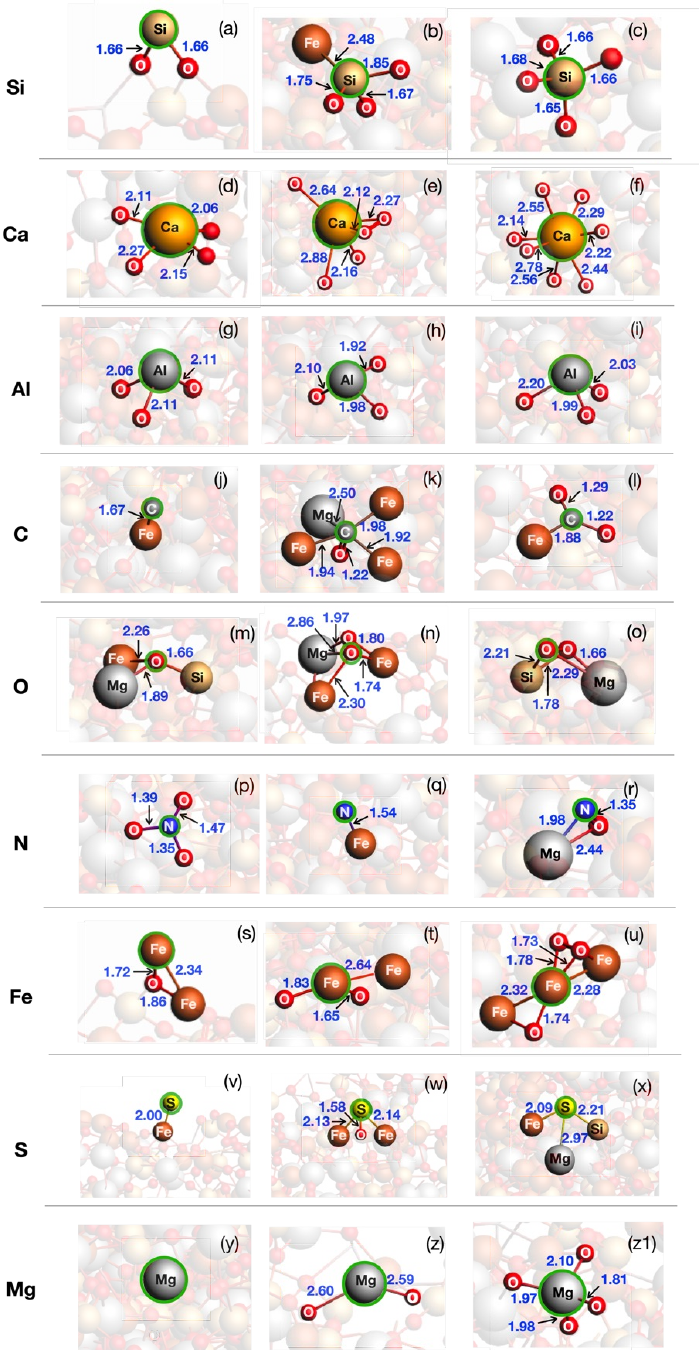}
      \caption{Selected optimized geometries of the local minima corresponding to the weakest, median, and strongest binding energies. The absorbed atom is indicated by a green circle.}
         \label{FigVibStab2}
   \end{figure}

\begin{table*}
\caption{\label{t7}Statistics of computed binding energy distributions for each element: mean, median, standard deviation, minimum, maximum, exponentiated logarithmic mean and exponentiated logarithmic standard deviation of the binding energies, as well as the Kolmogorov-Smirnov test p-value for statistical consistency with a lognormal distribution.}
\centering
\begin{tabular}{ccccccccc}
\hline
\hline
Atom & Mean (eV) & Median (eV) & St. Dev. (eV) & Min. (eV) & Max. (eV) & Log. Mean (eV) & Log St. Dev. (eV) & $p_{{\rm lognorm, KS}}$\\
\hline
Si & 15.31 & 14.76 & 5.52 & 10.40 & 22.21 & 15.14 & 1.16 & 0.537 \\
Ca & 13.52 & 12.66 & 7.83 & 10.28 & 22.28 & 13.27 & 1.21 & 0.015 \\
Al & 12.79 & 12.83 & 2.10 & 10.09 & 16.98 & 12.71 & 1.12 & 0.752 \\
C & 9.20 & 9.51 & 8.61 & 4.63 & 16.44 & 8.71 & 1.40 & 0.077 \\
O & 8.10 & 8.10 & 1.67 & 6.02 & 12.00 & 8.00 & 1.17 & 0.204 \\
N & 6.37 & 6.22 & 1.86 & 3.26 & 10.60 & 6.22 & 1.25 & 0.283 \\
Fe & 5.92 & 6.04 & 1.97 & 3.12 & 10.04 & 5.75 & 1.27 & 0.534 \\
S & 5.17 & 5.21 & 1.29 & 3.21 & 10.83 & 5.06 & 1.23 & 0.310 \\
Mg & 2.57 & 2.42 & 0.52 & 1.55 & 5.00 & 2.47 & 1.31 & 0.052 \\
\hline
\end{tabular}

\end{table*}

\begin{figure*}
\centering
\includegraphics[width=\hsize]{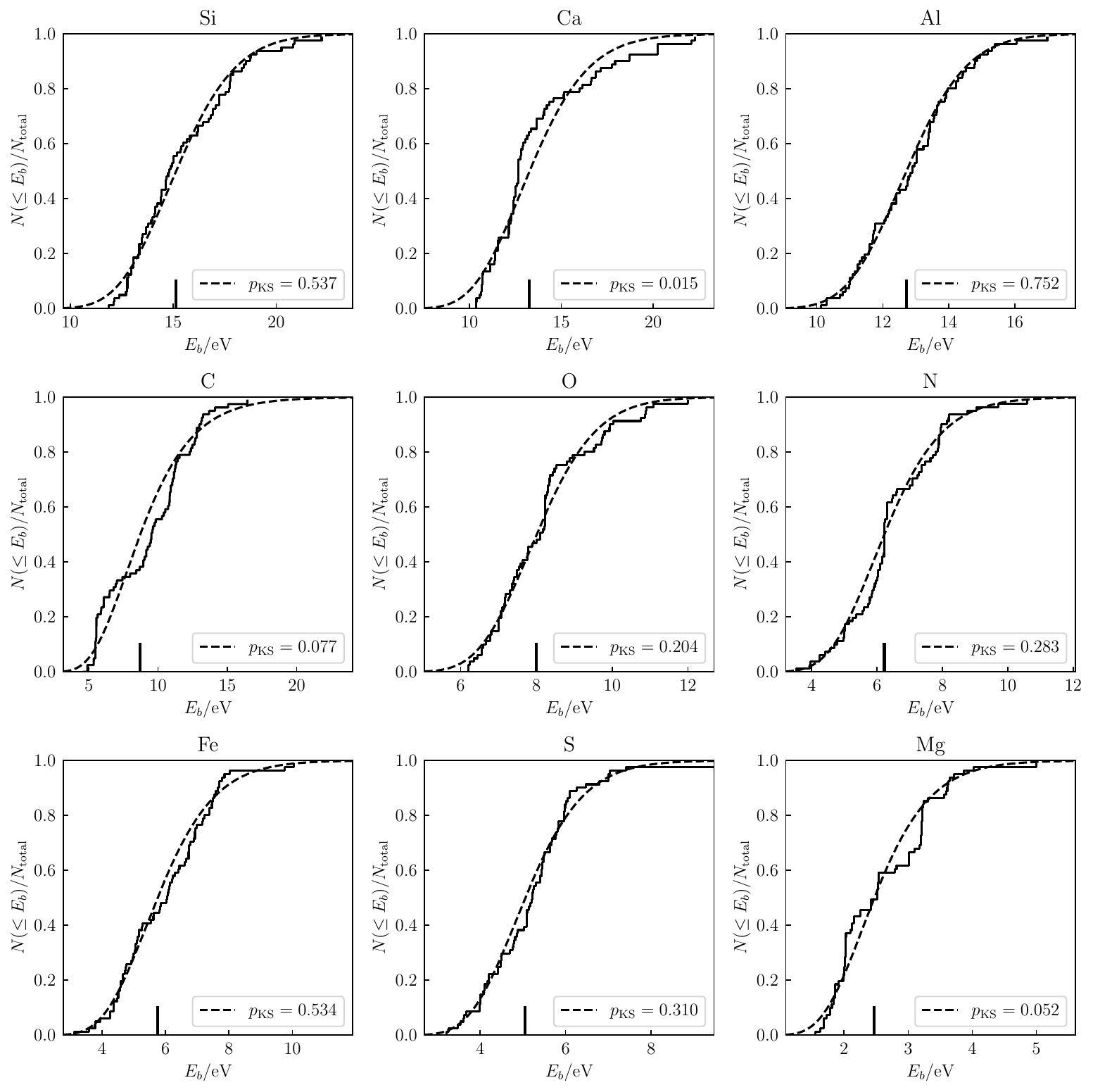}
  \caption{Cumulative Distribution Functions (CDFs) of binding energies of atoms on the FeMgSiO\textsubscript{4} dust particle model. Solid lines are the CDFs from the molecular dynamics calculations. Dashed lines are the CDFs of the best-fit log-normal distributions, labeled by the Kolmogorov-Smirnov test p-value. The vertical line indicates the exponentiated logarithmic mean.}
     \label{FigVibStab2_alt}
\end{figure*}

The C atom establishes chemical bonds with the Fe, Al, Si and O atoms on the FeMgSiO\textsubscript{4} dust particle surface. The computed binding energy of the C atom is the weakest (4.63 eV) when the C interacts with only one Fe (Figure 2j), and the binding energy is the strongest (16.44 eV) when the C atom interacts with two O and one Fe (Figure 2l)). The binding energy of the C atom on the surface is relatively weak compared to that of the Si atom, which sits in the same group of the periodic table. This is because C does not bind well into the silicate, while Si can form strong Si-O-Si bonds in the oxygen-rich environment of silicates. In the case of the O atom, the calculated binding energies are in the range of 6.02 – 12.00 eV, and the computed median binding energy is 8.10 eV. The geometries of the optimized structures suggested that the O atom can create chemical bonds with any atom on the surface (Figure 2m-o). The N atom shows relatively weak binding energy (3.3 - 10.6 eV, Figure 2p-r), and can interact with any atom on the particle surface.

The Fe atom makes bonds with O or Fe on the particle surface. The computed weakest binding energy of the Fe atom is 3.12 eV, where the Fe atom makes bonds with one surface O and one surface Fe (Figure 2s). When the Fe atom makes bonds with three surface O and two surface Fe, the binding energy is the strongest (10.04 eV, Figure 2u). In the case of the S atom, one to three chemical bonds can be made with any atom on the surface (Figure 2v-x), and the calculated binding energies range from 3.21 eV to 10.83 eV. Compared to the O atom, which is in the same group of the periodic table as S, the computed binding energy of the S atom is relatively weak due to differences in electronegativity. Moreover, the atomic radius of O is smaller than that of S, and therefore O makes stronger bonds with the silicate surface. The relatively large atomic radius of a sulfur atom results in weaker orbital overlap with the surface atoms, thereby forming weaker bonds compared to O atoms. The Mg atom either makes no chemical bond with the surface atoms of the FeMgSiO\textsubscript{4} dust particle or establishes two to four chemical bonds with the O atoms of the surface. In the former case, the binding energy is very low (e.g., 1.55 eV, Figure 2y), while the binding energy is the strongest when the Mg atom makes four chemical bonds with O on the particle surface (Figure 2Z1). 

\begin{table}
\caption{Interaction of adsorbed atoms with dust particle surface.}             
\label{table:1}      
\centering                          
\begin{tabular}{c c c}         
\hline\hline                 
Atom & Bonded & Non-bonded\\    
\hline                        
   Si & Fe, O & Mg, Si\\ 
   Ca & O & Fe, Mg, Si\\ 
   Al & O & Fe, Mg, Si\\ 
   C & Fe, Mg, Si, O & - \\ 
   O & Fe, Mg, Si, O & - \\ 
   N & Fe, Mg, Si, O & - \\
   Fe & Fe, O & Mg, Si\\ 
   S & Fe, Mg, Si, O & - \\ 
   Mg & O & Fe, Mg, Si\\ 
\hline                                   
\end{tabular}
\end{table}

Visual inspection of the computed structures confirmed that the C, O, N, and S atoms can form bonds with any atom of the dust particle after adsorption (Table 2). The Si and Fe atoms can form bonds only with Fe and O atoms of the dust particle, whereas Ca, Al, and Mg atoms can form bonds only with O atoms of the surface upon adsorption.     
 
The computed probability distributions of binding energies of atoms on the FeMgSiO\textsubscript{4} dust particle model are shown in Figure 3, which for all but one element (Ca) are statistically consistent with a log-normal distribution as evaluated by a Kolmogorov-Smirnov (KS) test, implemented in the scipy.stats Python library \citep{2020SciPy-NMeth}. This is perhaps surprising; one could imagine that the finite number of bond types possible for each element would result in characteristic values of the binding energy distribution that would make it more complicated. However, it appears that the amorphous nature of the grain model used in the calculations results in a sufficient variety of bonding configurations for each element that the central limit theorem holds approximately. This may also reflect the relative insensitivity of the KS test to distribution details, or to the small number of binding sites in our sample.

\subsection{Sublimation Temperatures}

\begin{figure}
    \centering
    \includegraphics[width=\hsize]{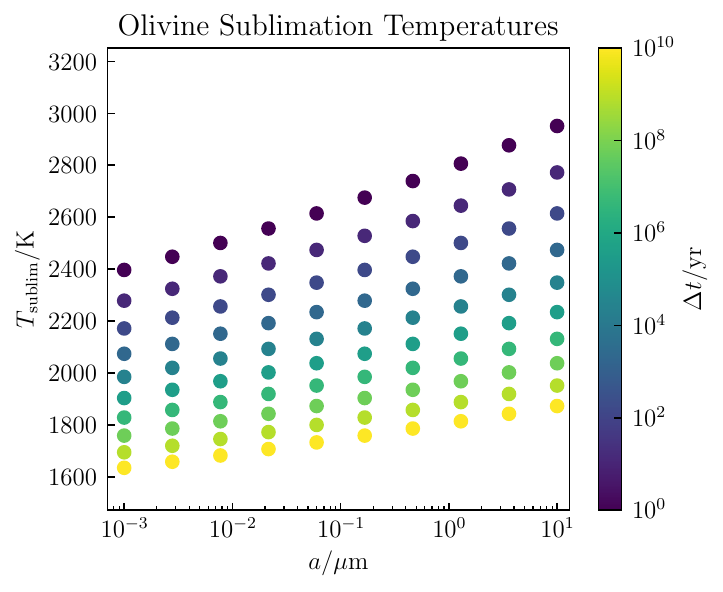}
    \caption{Sublimation temperatures for silicate dust calculated from binding energy distributions.}
    \label{fig:sublimation_temps}
\end{figure}

The binding energies computed in this study determine the rate of thermal desorption -- i.e. sublimation -- of atoms from the grain surface. To predict grain sublimation temperatures, we begin with the mass flux of grains due to thermal desorption from the surface given by the Polanyi-Wigner equation \citep{Polanyi_1928,Potapov_2018} 

\begin{equation}
\frac{dm_g}{dt} = -\sum_i m_i N_i \nu_{0,i}\exp\left(\frac{-U_i}{k_{\rm B}T}\right)
\end{equation}

\noindent where $i$ indexes all the elements of which the grain is composed, $N_i$ is the number of each element on the surface, $\nu_{0,i}$ is the characteristic vibration frequency of each element, $U_i$ is the surface binding energy of each element -- here assumed to be constant for all atoms of the same element -- and $T$ is the dust grain temperature. We can express the number of atoms of each element on the surface as 

\begin{equation}
N_i = \frac{f_i v_{\rm mono}\rho_g }{m_i}
\end{equation}

\noindent where $\rho_g$ is the grain material density, $f_i$ is the mass fraction of each constituent element given by 

\begin{equation}
f_i = \frac{\sum_i m_i f_{N,i}}{\sum_i m_i}
\end{equation}

\noindent and $v_{\rm mono}$, the volume of the surface ``monolayer'' -- the layer with atoms capable of escaping from the surface -- is 

\begin{equation}
v_{\rm mono} \approx 4\pi a^2 \langle x\rangle \ , 
\end{equation}

\noindent where $\langle x\rangle$ is the average monolayer thickness  (assumed $<< a$) defined by

\begin{equation}
\langle x\rangle = \sum_i N_{{\rm mono},i}x_i \ , 
\end{equation}

\noindent where $N_{{\rm mono},i}$ is the number of elements $i$ in each monolayer ``unit'', each with a characteristic size $x_i$. For the assumed chemical composition of our silicate grain, $N_{\rm mono, Fe} = N_{\rm mono, Mg} = N_{\rm mono, Si} = 1$ and $N_{\rm mono, O} = 4$. Note that we can relate this to the grain material density $\rho_g$ as follows

\begin{equation}
\rho_g = \frac{\sum_i N_{{\rm mono},i}m_i}{(\sum_i N_{{\rm mono},i}x_i)^3} \implies \sum_i N_{{\rm mono},i}x_i = \left(\frac{\sum_i N_{{\rm mono},i}m_i}{\rho_g}\right)^{1/3}
\end{equation}

\noindent giving 

\begin{equation}
\frac{dm_g}{dt} = -4\pi a^2\rho_g\left(\frac{\sum_i N_{{\rm mono},i}m_i}{\rho_g}\right)^{1/3}\left[\sum_i f_i\nu_{0,i}\exp\left(\frac{-U_i}{k_{\rm B}T}\right)\right].
\end{equation}

\noindent This mass evolution equation can be re-expressed in terms of rate of change of grain effective radius $a$:

\begin{equation}
m = \frac{4}{3}\pi a^3\rho_g \implies \frac{dm_g}{dt} = 4\pi a^2\rho_g \frac{da}{dt}
\end{equation}

\noindent we note that the above assumes assumes constant $\rho_g$, which is incorrect in principle because more weakly bound elements will desorb more quickly, changing the grain composition on the surface, but we ignore this effect because it is negligible for grains with sizes much larger than a single monolayer. Our grain radius evolution equation is then

\begin{equation}
\frac{da}{dt} = -\left(\frac{\sum_i N_{{\rm mono},i}m_i}{\rho_g}\right)^{1/3}\left[\sum_i f_i\nu_{0,i}\exp\left(\frac{-U_i}{k_{\rm B}T}\right)\right].
\end{equation}

However, this expression assumes that all atoms of a given element $i$ have the same surface binding energy  $U_i$, which we predict from our calculations is not strictly correct. To account for a probability distribution function of binding energies for a given atom $P(U_i)$, we replace the Boltzmann factor by its expectation value over this distribution:

\begin{equation}
e^{-U/k_{\rm B}T} \to \int_0^{\infty}P(U)e^{-U/k_{\rm B}T}dU \equiv \left\langle\exp\left(\frac{-U_i}{k_{\rm B}T}\right)\right\rangle
\end{equation}

\noindent giving us finally

\begin{equation}\label{eq:grain_radius_rate}
\frac{da}{dt} = -\left(\frac{\sum_i N_{{\rm mono},i}m_i}{\rho_g}\right)^{1/3}\left[\sum_i f_i\nu_{0,i}\left\langle\exp\left(\frac{-U_i}{k_{\rm B}T}\right)\right\rangle\right].
\end{equation}

However, this equation only holds for desorption of the most loosely bound atoms in the first monolayer: the sublimation thereafter will be rate-limited by the most strongly bound element, in our case Si. Therefore, the long-term grain radius evolution differential equation will be

\begin{equation}\label{eq:grain_radius_rate_limited}
\frac{da}{dt} = -\left(\frac{\sum_i N_{{\rm mono},i}m_i}{\rho_g}\right)^{1/3}\nu_{0,{\rm Si}}\left\langle\exp\left(\frac{-U_{\rm Si}}{k_{\rm B}T}\right)\right\rangle.
\end{equation}

To estimate grain lifetimes $\Delta t$ or equivalently sublimation temperatures $T_{\rm sublim}$, assuming $T$ in eq. \ref{eq:grain_radius_rate} does not evolve with time or grain radius allows this differential equation to be integrated trivially to obtain

\begin{equation}
\Delta t_{\rm subl}(a, T) = a \left(\frac{\sum_i N_{{\rm mono},i}m_i}{\rho_g}\right)^{-1/3}\left[\nu_{0,{\rm Si}}\left\langle\exp\left(\frac{-U_{\rm Si}}{k_{\rm B}T}\right)\right\rangle\right]^{-1}
\end{equation}

\noindent for the grain lifetime $\Delta t_{\rm subl}$ as a function of grain radius and temperature, or alternatively define the sublimation temperature $T_{\rm subl}(a | \Delta t)$ at a given grain radius and lifetime by

\begin{equation}
\left[\nu_{0,{\rm Si}}\left\langle\exp\left(\frac{-U_{\rm Si}}{k_{\rm B}T_{\rm subl}(a | \Delta t)}\right)\right\rangle\right]= \frac{a}{\Delta t} \left(\frac{\sum_i N_{{\rm mono},i}m_i}{\rho_g}\right)^{-1/3}.
\end{equation}

\noindent While $T_{\rm subl}(a | \Delta t)$ precludes a symbolic expression, this equation can be solved for $T$ numerically.

The predicted values of $T_{\rm subl}(a | \Delta t)$ for the relevant range of $a$ and $\Delta t$ are shown in Fig. \ref{fig:sublimation_temps}. In this calculation we assumed a temperature-dependent vibrational frequency for silicon bonds described in Appendix \ref{appendix:freq}. Because of this, we subtract a zero-point energy of 0.2eV from the Si binding energy distribution. We also adopted a grain material density of $\rho_g = 3.32{\rm g}\;{\rm cm}^{-3}$ for olivine\footnote{\url{https://webmineral.com/data/Olivine.shtml}}. To estimate the probability distribution function $P_i(U)$ for each element from the molecular dynamics calculations we employed a Kernel Density Estimate (KDE) with the Epanechnikov kernel of constant size set to the optimal value using Silverman's ``rule-of-thumb'' \citep{Silverman1986}. Sublimation temperatures range from $\approx 1600{\rm K}$ to $\approx 3000{\rm K}$ depending on assumed grain size and lifetime, and are approximately consistent with previous estimates in the literature \cite{Pollack1994, Waxman2000} as well as temperatures of dust observed in AGN environments \citep{Barvainis1987, Kishimoto2007}. We note that in principle, the compositional changes due to different thermal desorption rates for different elements will likely change the binding energy distribution as the desorption goes on. However, we expect that these changes will not be so large as to qualitatively change our conclusions.  This is partly helped by the  grain being amorphous  which should facilitate the rearrangement of bonds so that they remain statistically similarly distributed as the grain sublimates.

\section{Discussion}\label{sec:discussion}

 The sublimation temperatures shown in Fig. \ref{fig:sublimation_temps} are much higher than typical temperatures of dust in the ambient ISM, confirming the standard assumption that thermal desorption is generally unimportant for dust grain evolution in these environments, and these grains are stable to thermal sublimation on cosmological timescales. However, these values are comparable to or much higher than the observed or predicted temperatures of dust in the most extreme interstellar environments -- asymptotic giant branch (AGB) stars \citep[e.g.][]{Lorenz-Martins2000, Suh2004, Lagadec2005, HerasHony2005, Groenewegen2009, Goldman2017, Hofner2022}, supernova remnants \citep[SNRs, see][]{Micelotta2018}, and the dusty tori of active galactic nuclei \citep[AGN, e.g.][]{Barvainis1987, Oybau2011,Dexter2020,Nenkova2008,Netzer2015,Weigelt2012,Sakamoto2021}. Therefore, these values will enable the precise calculation of grain growth and destruction rates in the full variety of interstellar environments that they experience.

 The binding energies of carbon atoms are of particular interest because interstellar dust is often assumed to consist of two separate chemical species, one which is primarily (hydro)carbonaceous and the other being silicate as is the focus of this study \cite[see][and references therein]{Draine2011}. The ability of carbon to bond to the silicate substrate, as predicted by our calculations, challenges this canonical chemical dichotomy. However, a full assessment of the ability of carbon atoms to bond to and participate to the growth of silicate dust will require dynamical calculations of sticking coefficients, which we will present in a forthcoming work. 

Similarly, the binding energies of nitrogen also suggest that it can, in principle, stick to amorphous silicate dust grains. This is possibly in contradiction with the observational result that nitrogen is not strongly depleted from the gas phase of the ISM \citep{Jenkins2009}. However, depletion measurements have large uncertainties, and this result may require the revision of existing dust models \cite[e.g.][]{jones2013evolution, HensleyDraine2023} to account for the possibility of nitrogen inclusion in candidate grain materials.

\section{Conclusions}\label{sec:conclusion}

The binding energy of C, O, Mg, Si, S, Al, Ca, Fe, and Ni atoms on a FeMgSiO\textsubscript{4} periodic dust particle model surface were computed using the GNF1-xTB method. The GNF1-xTB approach, a semi-empirical tight-binding method, has been shown to provide reliable and computationally efficient results for complex molecular systems. Computed data showed a range of binding energies for each atom. The median of the binding energies follows the order of Si (14.8 eV) > Al (12.8) > Ca (12.7) > C (9.5) > O (8.1 eV) > N (6.2 eV) > Fe (6.0 eV) > S (5.2 eV) > Mg (2.4 eV). At the present level of theory, the Si and Ca atoms can react with the surface, giving rise to several chemical bonds between the atom (Si or Ca) and the particle surface atoms. Thus, the Si and Ca atoms penetrate in the FeMgSiO\textsubscript{4} dust particle and stay near the particle surface (i.e., they do not stay on the surface). Other atoms adsorbed on the FeMgSiO\textsubscript{4} dust particle and sit on the particle surface. We found near-log-normal binding energy distributions for most elements, and used these distributions to estimate sublimation temperatures for silicate dust grains which are consistent with observations and previous theoretical estimates. These unprecedentedly precise characterizations of silicate dust surface chemical properties will enable accurate calculations of dust processes in the full diversity of ISM environments, enabling sophisticated predictions about the interplay between ISM dynamics and dust evolution. 

\begin{acknowledgements}
     We acknowledge support from the Knut and Alice Wallenberg Foundation under grant no. KAW 2020.0081. Supercomputing resources provided by Chalmers e-Commons at Chalmers and the National Academic Infrastructure for Supercomputing in Sweden (NAISS), partially funded by the Swedish Research Council through grant agreement no. 2022-06725. We thank Theo Khouri for providing references on AGB star dust temperatures.
\end{acknowledgements}

%-------------------------------------------------------------------
\bibliographystyle{aa}
\bibliography{references}

%-------------------------------------------------------------------
\begin{appendix}\label{appendix:MD_sim} %First online appendix
\section{}
In Fig A.1. the MD simulation results show a typical equilibration behavior of the system. Initially both the total energy and the temperature increase sharply, suggesting that the system is relaxing from the initial configuration. As the simulation progresses, the total energy gradually stabilizes, while the temperature oscillates around 5000 K.
\begin{figure}
\centering

   \includegraphics[width=\hsize]{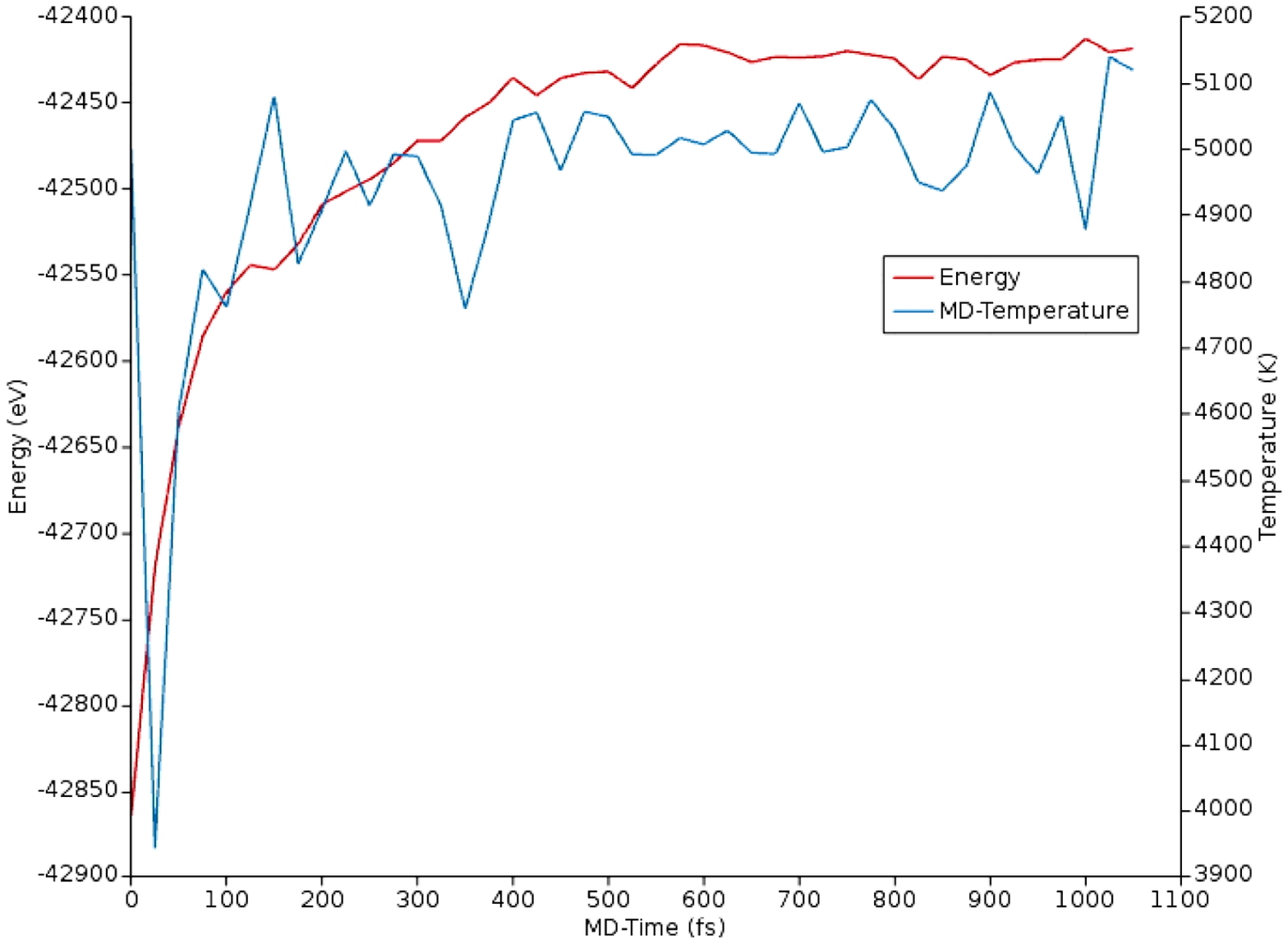}
      \caption{Total energy and temperature profiles during MD simulation to amorphize molecular structure. }
         
   \end{figure}

\begin{sidewaystable*}
\caption{Energy (AU) of the optimized local minima}\label{table:data}
\centering
\begin{tabular}{ccccccccccc}
\hline
Local minima & Si & Ca & Al & C & O & N & Fe & S & Mg \\
\hline
1&	-1587,57197	&	-1586,43520	&	-1586,94069	&	-1587,68063	&	-1590,05078	&	-1588,47314	&	-1588,48376	&	-1589,21380	&	-1586,26415 \\ 
2&	-1587,57408	&	-1586,51035	&	-1586,94619	&	-1587,41236	&	-1590,07185	&	-1588,49840	&	-1588,49295	&	-1589,12350	&	-1586,27012 \\
3&	-1587,57394	&	-1586,51295	&	-1587,01553	&	-1587,55914	&	-1590,02947	&	-1588,61208	&	-1588,49726	&	-1589,17549	&	-1586,30910 \\
4&	-1587,85320	&	-1586,51297	&	-1587,01599	&	-1587,43612	&	-1590,07881	&	-1588,56633	&	-1588,49308	&	-1589,20361	&	-1586,30917 \\
5&	-1587,66498	&	-1586,51288	&	-1587,01554	&	-1587,44646	&	-1590,06885	&	-1588,57209	&	-1588,48616	&	-1589,19951	&	-1586,30905 \\
6&	-1587,51906	&	-1586,51287	&	-1587,01561	&	-1587,66274	&	-1590,11872	&	-1588,74516	&	-1588,49612	&	-1589,15286	&	-1586,25808 \\
7&	-1587,63315	&	-1586,53064	&	-1587,00357	&	-1587,65334	&	-1590,11516	&	-1588,64517	&	-1588,46946	&	-1589,20583	&	-1586,25232 \\
8&	-1587,77407	&	-1586,56464	&	-1586,93288	&	-1587,62225	&	-1590,17463	&	-1588,52945	&	-1588,48852	&	-1589,15284	&	-1586,25808 \\
9&	-1587,55254	&	-1586,50760	&	-1586,95394	&	-1587,71753	&	-1590,04917	&	-1588,50977	&	-1588,53231	&	-1589,17836	&	-1586,28358 \\
10&	-1587,55255	&	-1586,60454	&	-1586,94931	&	-1587,58038	&	-1590,16946	&	-1588,58353	&	-1588,45917	&	-1589,17830	&	-1586,26081 \\
11&	-1587,53765	&	-1586,52136	&	-1586,94909	&	-1587,37626	&	-1590,07134	&	-1588,53831	&	-1588,58229	&	-1589,10841	&	-1586,26413 \\
12&	-1587,62155	&	-1586,51307	&	-1586,99655	&	-1587,67536	&	-1590,09900	&	-1588,62520	&	-1588,47149	&	-1589,16328	&	-1586,30943 \\
13&	-1587,65296	&	-1586,51855	&	-1586,92955	&	-1587,45528	&	-1590,11327	&	-1588,64112	&	-1588,45923	&	-1589,18610	&	-1586,30995 \\
14&	-1587,59729	&	-1586,51295	&	-1586,95135	&	-1587,65702	&	-1590,08769	&	-1588,48353	&	-1588,41727	&	-1589,16951	&	-1586,31264 \\
15&	-1587,52730	&	-1586,50313	&	-1586,91442	&	-1587,51599	&	-1590,04153	&	-1588,58050	&	-1588,41716	&	-1589,14571	&	-1586,26478 \\
16&	-1587,67907	&	-1586,51847	&	-1586,91445	&	-1587,49040	&	-1590,20923	&	-1588,50964	&	-1588,46343	&	-1589,16663	&	-1586,24756 \\
17&	-1587,57969	&	-1586,60491	&	-1586,93618	&	-1587,52589	&	-1590,16097	&	-1588,51602	&	-1588,46879	&	-1589,11020	&	-1586,24766 \\
18&	-1587,59548	&	-1586,56503	&	-1587,10361	&	-1587,53377	&	-1590,05867	&	-1588,53834	&	-1588,48881	&	-1589,16829	&	-1586,28353 \\
19&	-1587,55946	&	-1586,63673	&	-1587,10015	&	-1587,53381	&	-1590,25111	&	-1588,62950	&	-1588,59915	&	-1589,17823	&	-1586,25250 \\
20&	-1587,55919	&	-1586,49354	&	-1587,02863	&	-1587,37626	&	-1590,08387	&	-1588,64395	&	-1588,50735	&	-1589,10832	&	-1586,28352 \\
21&	-1587,54976	&	-1586,66949	&	-1587,00532	&	-1587,57577	&	-1590,08394	&	-1588,64401	&	-1588,49066	&	-1589,39089	&	-1586,30797 \\
22&	-1587,55252	&	-1586,57526	&	-1586,96750	&	-1587,59189	&	-1590,11063	&	-1588,56888	&	-1588,46776	&	-1589,18113	&	-1586,25613 \\
23&	-1587,70270	&	-1586,50309	&	-1587,01058	&	-1587,66884	&	-1590,13844	&	-1588,67764	&	-1588,59274	&	-1589,16729	&	-1586,25900 \\
24&	-1587,46423	&	-1586,49638	&	-1587,02335	&	-1587,66882	&	-1590,11399	&	-1588,65654	&	-1588,55846	&	-1589,20494	&	-1586,25908 \\
25&	-1587,75822	&	-1586,70809	&	-1586,99238	&	-1587,66498	&	-1590,17439	&	-1588,52653	&	-1588,50977	&	-1589,21545	&	-1586,26400 \\
26&	-1587,67905	&	-1586,73681	&	-1586,96213	&	-1587,55198	&	-1590,21039	&	-1588,53743	&	-1588,46348	&	-1589,14066	&	-1586,25050 \\
27&	-1587,62170	&	-1586,67851	&	-1586,96240	&	-1587,52920	&	-1590,04491	&	-1588,53746	&	-1588,46346	&	-1589,19987	&	-1586,25436 \\
28&	-1587,60229	&	-1586,49350	&	-1587,02870	&	-1587,80413	&	-1590,06556	&	-1588,56437	&	-1588,51676	&	-1589,20337	&	-1586,26380 \\
29&	-1587,90180	&	-1586,49334	&	-1587,02894	&	-1587,41975	&	-1590,06553	&	-1588,63531	&	-1588,55182	&	-1589,14189	&	-1586,26486 \\
30&	-1587,85049	&	-1586,65882	&	-1587,08660	&	-1587,42023	&	-1590,10893	&	-1588,57572	&	-1588,59007	&	-1589,15613	&	-1586,37527 \\
31&	-1587,60271	&	-1586,46618	&	-1586,98353	&	-1587,41996	&	-1590,08914	&	-1588,56967	&	-1588,51005	&	-1589,15616	&	-1586,26472 \\
32&	-1587,61288	&	-1586,51736	&	-1587,01117	&	-1587,54294	&	-1590,04939	&	-1588,65587	&	-1588,44760	&	-1589,15606	&	-1586,25914 \\
33&	-1587,56276	&	-1586,46504	&	-1586,98737	&	-1587,68584	&	-1590,16782	&	-1588,59731	&	-1588,55853	&	-1589,19139	&	-1586,25918 \\
34&	-1587,73739	&	-1586,50292	&	-1587,04011	&	-1587,59553	&	-1590,06586	&	-1588,71220	&	-1588,55694	&	-1589,26418	&	-1586,26476 \\
35&	-1587,56268	&	-1586,66882	&	-1586,98714	&	-1587,52550	&	-1590,11592	&	-1588,58626	&	-1588,55867	&	-1589,14155	&	-1586,26472 \\
36&	-1587,78476	&	-1586,73681	&	-1586,95712	&	-1587,60796	&	-1590,20469	&	-1588,57470	&	-1588,47421	&	-1589,12536	&	-1586,26791 \\
37&	-1587,55132	&	-1586,79424	&	-1586,96680	&	-1587,58447	&	-1590,06582	&	-1588,58620	&	-1588,47747	&	-1589,19822	&	-1586,27978 \\
38&	-1587,62811	&	-1586,79425	&	-1587,09545	&	-1587,66925	&	-1590,06832	&	-1588,57091	&	-1588,55101	&	-1589,17041	&	-1586,33918 \\
39&	-1587,63085	&	-1586,49344	&	-1587,12790	&	-1587,75252	&	-1590,08602	&	-1588,68798	&	-1588,67352	&	-1589,22159	&	-1586,26471 \\
40&	-1587,55468	&	-1586,70162	&	-1587,08327	&	-1587,61011	&	-1590,10298	&	-1588,64686	&	-1588,56963	&	-1589,24943	&	-1586,30178 \\
\hline
\end{tabular}
\end{sidewaystable*}

\begin{sidewaystable*}
\caption{Energy (AU) of the optimized local minima ( Continue....)}\label{table:data2}
\centering
\begin{tabular}{ccccccccccc}
\hline
Local minima & Si & Ca & Al & C & O & N & Fe & S & Mg \\
\hline
41	&	-1587,67768	&	-1586,58395	&	-1587,01326	&	-1587,70182	&	-1590,15897	&	-1588,64689	&	-1588,55097	&	-1589,18998	&	-1586,26860 \\
42	&	-1587,56317	&	-1586,53454	&	-1586,94418	&	-1587,43518	&	-1590,16594	&	-1588,62185	&	-1588,54635	&	-1589,21148	&	-1586,26853 \\
43	&	-1587,58595	&	-1586,86235	&	-1587,03447	&	-1587,45210	&	-1590,16715	&	-1588,64802	&	-1588,44238	&	-1589,21263	&	-1586,26509 \\
44	&	-1587,70649	&	-1586,86944	&	-1587,16234	&	-1587,61762	&	-1590,06832	&	-1588,61517	&	-1588,54034	&	-1589,17805	&	-1586,26520 \\
45	&	-1587,61356	&	-1586,79417	&	-1586,96686	&	-1587,60312	&	-1590,12034	&	-1588,63819	&	-1588,57820	&	-1589,24899	&	-1586,26481 \\
46	&	-1587,73740	&	-1586,54915	&	-1587,03835	&	-1587,39943	&	-1590,17814	&	-1588,58285	&	-1588,43512	&	-1589,24103	&	-1586,27986 \\
47	&	-1587,76211	&	-1586,43809	&	-1586,97955	&	-1587,40020	&	-1590,12227	&	-1588,61534	&	-1588,58691	&	-1589,15552	&	-1586,30944 \\
48	&	-1587,63593	&	-1586,46548	&	-1586,96286	&	-1587,61351	&	-1590,07200	&	-1588,64495	&	-1588,66297	&	-1589,14541	&	-1586,30935 \\
49	&	-1587,68839	&	-1586,46642	&	-1587,04944	&	-1587,39312	&	-1590,08087	&	-1588,56014	&	-1588,53856	&	-1589,21556	&	-1586,30512 \\
50	&	-1587,62258	&	-1586,57037	&	-1587,07070	&	-1587,39322	&	-1590,05755	&	-1588,55095	&	-1588,55288	&	-1589,12062	&	-1586,30178 \\
51	&	-1587,62175	&	-1586,47348	&	-1587,08159	&	-1587,68242	&	-1590,13520	&	-1588,58272	&	-1588,58502	&	-1589,18427	&	-1586,30179 \\
52	&	-1587,64139	&	-1586,50276	&	-1587,03858	&	-1587,40048	&	-1590,11116	&	-1588,59196	&	-1588,51748	&	-1589,18995	&	-1586,30174 \\
53	&	-1587,73667	&	-1586,54922	&	-1586,98058	&	-1587,40041	&	-1590,11112	&	-1588,58289	&	-1588,56672	&	-1589,21073	&	-1586,32303 \\
54	&	-1587,72862	&	-1586,66120	&	-1587,06827	&	-1587,60437	&	-1590,11117	&	-1588,58371	&	-1588,57830	&	-1589,19096	&	-1586,32522 \\
55	&	-1587,58594	&	-1586,53839	&	-1586,96869	&	-1587,54734	&	-1590,11150	&	-1588,58290	&	-1588,51888	&	-1589,19370	&	-1586,28436 \\
56	&	-1587,65969	&	-1586,42464	&	-1586,96929	&	-1587,53752	&	-1590,03764	&	-1588,57558	&	-1588,49116	&	-1589,18294	&	-1586,32430 \\
57	&	-1587,76614	&	-1586,44075	&	-1587,03825	&	-1587,59916	&	-1590,11584	&	-1588,58603	&	-1588,52831	&	-1589,22977	&	-1586,32372 \\
58	&	-1587,64798	&	-1586,52701	&	-1587,02804	&	-1587,64990	&	-1590,09371	&	-1588,65376	&	-1588,54808	&	-1589,20544	&	-1586,30930 \\
59	&	-1587,74350	&	-1586,50715	&	-1587,04731	&	-1587,61183	&	-1590,11141	&	-1588,58332	&	-1588,53164	&	-1589,25049	&	-1586,27361 \\
60	&	-1587,61362	&	-1586,44112	&	-1587,05055	&	-1587,59685	&	-1590,05354	&	-1588,62510	&	-1588,58054	&	-1589,16623	&	-1586,30856 \\
61	&	-1587,61224	&	-1586,42730	&	-1587,06222	&	-1587,59642	&	-1590,13275	&	-1588,58080	&	-1588,56765	&	-1589,19166	&	-1586,28401 \\
62	&	-1587,71759	&	-1586,42801	&	-1587,05989	&	-1587,40045	&	-1590,11116	&	-1588,59627	&	-1588,57943	&	-1589,21083	&	-1586,28412 \\
63	&	-1587,73540	&	-1586,54920	&	-1586,96614	&	-1587,40050	&	-1590,11118	&	-1588,58290	&	-1588,53481	&	-1589,19276	&	-1586,32208 \\
64	&	-1587,70683	&	-1586,47217	&	-1586,99246	&	-1587,54558	&	-1590,20947	&	-1588,54718	&	-1588,52540	&	-1589,19125	&	-1586,32457 \\
65	&	-1587,64949	&	-1586,43829	&	-1587,00646	&	-1587,50690	&	-1590,07669	&	-1588,56684	&	-1588,47213	&	-1589,18367	&	-1586,25714 \\
66	&	-1587,57864	&	-1586,47123	&	-1586,90826	&	-1587,39765	&	-1590,09460	&	-1588,58608	&	-1588,47792	&	-1589,13772	&	-1586,27343 \\
67	&	-1587,77367	&	-1586,45530	&	-1587,03046	&	-1587,39764	&	-1590,11367	&	-1588,59277	&	-1588,52638	&	-1589,13730	&	-1586,33338 \\
68	&	-1587,60640	&	-1586,50731	&	-1587,04927	&	-1587,39777	&	-1590,09369	&	-1588,58599	&	-1588,52921	&	-1589,14390	&	-1586,32759 \\
69	&	-1587,60149	&	-1586,45526	&	-1587,03058	&	-1587,55030	&	-1590,14990	&	-1588,49916	&	-1588,52708	&	-1589,17996	&	-1586,30853 \\
70	&	-1587,71134	&	-1586,44120	&	-1587,00510	&	-1587,59608	&	-1590,11119	&	-1588,58305	&	-1588,55142	&	-1589,19291	&	-1586,28424 \\
71	&	-1587,71799	&	-1586,44113	&	-1587,05983	&	-1587,41342	&	-1590,07256	&	-1588,58284	&	-1588,57495	&	-1589,21025	&	-1586,28380 \\
72	&	-1587,59033	&	-1586,49927	&	-1586,99255	&	-1587,40049	&	-1590,11108	&	-1588,58301	&	-1588,48796	&	-1589,21125	&	-1586,29363 \\
73	&	-1587,63524	&	-1586,52272	&	-1586,93959	&	-1587,54769	&	-1590,21707	&	-1588,52169	&	-1588,47417	&	-1589,12347	&	-1586,27019 \\
74	&	-1587,57405	&	-1586,43944	&	-1586,94069	&	-1587,55608	&	-1590,08020	&	-1588,62086	&	-1588,47686	&	-1589,13800	&	-1586,26475 \\
75	&	-1587,57870	&	-1586,47227	&	-1587,03503	&	-1587,36629	&	-1590,07902	&	-1588,64506	&	-1588,44179	&	-1589,11860	&	-1586,25251 \\
76	&	-1587,52960	&	-1586,50458	&	-1586,90680	&	-1587,39763	&	-1590,09462	&	-1588,57336	&	-1588,47047	&	-1589,13727	&	-1586,30875 \\
77	&	-1587,73630	&	-1586,64202	&	-1586,98532	&	-1587,39772	&	-1590,10647	&	-1588,57852	&	-1588,48574	&	-1589,13798	&	-1586,30899 \\
78	&	-1587,69804	&	-1586,51306	&	-1587,03074	&	-1587,40385	&	-1590,10655	&	-1588,57844	&	-1588,52624	&	-1589,18284	&	-1586,27961 \\
79	&	-1587,82985	&	-1586,42842	&	-1587,06710	&	-1587,47003	&	-1590,03586	&	-1588,55832	&	-1588,58661	&	-1589,18252	&	-1586,25603 \\
80	&	-1587,71776	&	-1586,56272	&	-1587,08041	&	-1587,59949	&	-1590,03600	&	-1588,53705	&	-1588,56071	&	-1589,17799	&	-1586,29439 \\
81	&	-1587,62591	&	-1586,49927	&	-1587,00841	&	-1587,60764	&	-1590,05755	&	-1588,54294	&	-1588,57550	&	-1589,19853	&	-1586,31017 \\
\hline
\end{tabular}
\end{sidewaystable*}

\section{}\label{appendix:freq}
To calculate estimates of the frequency factor (preexponential factor) for the desorption rate coefficient expression we have employed Transition State Theory (TST) \citep{Pitt_1994,Tait_2005}. The system is divided into the atom that is desorbing and the rest of the grain. The basic expression for the desorption rate constant is

\begin{equation}
    k(T)=\frac{k_{\rm B}T}{h}\frac{Q^{\ddag}}{Q^R}\exp\left(-\frac{E_{des}}{k_{\rm B}T}\right)
\end{equation}

\noindent where $E_{des}$ is the desorption energy from a binding site. The temperature, $T$, is the temperature of the grain surface. If there is no additional energy barrier to desorption $E_{des}=E_{bind}$. $Q^R$ and $Q^{\ddag}$ are the partition functions of the reactant state and transition state, respectively. With this definition the pre-exponential factor can be identified as the frequency factor in the Polanyi-Wigner expression. It should be borne in mind that the TST expression is in general an upper bound of the real rate constant if all interactions are treated with the highest accuracy possible.

The definition of reactant state and transition state is not unambiguous. For a strongly bonded atom at a surface it is natural to take the reactant state to be represented by the atom localized at a specific location and its motion in 3 degrees of freedom will be characterized by 3 vibrational frequencies. As the atom gains more kinetic energy its motion could eventually be characterized as 2-dimensional (hindered) translation parallel to the surface and 1-dimensional vibration in the direction normal to the surface. The transition state can be taken to be a dividing surface at a distance from the surface where the atom-surface interaction is insignificant. At that point motion normal to the surface is now translation in 1D that characterizes motion through the dividing surface, i.e., effectively desorption. 

The transition state is characterized by translational motion in 2D, and its extent parallel to the surface is usually taken to be the inverse of the surface binding site density, i.e., an effective area of the binding site. The atom at the transition state is effectively in the gas phase and its electronic degrees of freedom are therefore characterized by its electronic fine-structure states. For the adsorbed atom, the nature of its electronic state is not as clear since its electronic structure is not independent from the other atoms in the surface and it is not in a spherically symmetric environment. A choice can be made to assign its electronic spin multiplicity as the number of electronic states at the surface. The spin multiplicity for the single atom is however not necessarily well-characterized, but lacking more detailed information on the electronic structure it seems like a reasonable approximation.

Given the above information the partition function for the transition state in the expression for the rate constant can be expressed as
\begin{equation}
Q^{\ddag} = q_{trans,2D}q_{el}
\end{equation}

\noindent where 
\begin{equation}
    q_{trans,2D}=\frac{2\pi m k_{\rm B}TA}{h^2}
\end{equation}

\noindent where $m$ is the mass of the atom and $A$ is the typical area of the binding site as discussed above. The electronic partition function is expressed as 

\begin{equation}
    q_{el}=\sum_{i=1}^N{g_i\exp\left(-\frac{E_i}{k_{\rm B}T}\right)}
\end{equation}

\noindent where the sum covers all available electronic fine-structure (spin-orbit) states $i$ of the atom with energies $E_i$ and degeneracy $g_i$.

The corresponding partition function for the reactant state is expressed as

\begin{equation}
    Q^R=q_{vib}q_{el}
\end{equation}

\noindent where the vibrational partition function is taken to be that of the collection of harmonic oscillators describing the three vibrational degrees of freedom of the atom at the surface is:

\begin{equation}
    q_{vib}=\prod_{i=1}^3\frac{1}{1-\exp(E_i/k_{\rm B}T)}
\end{equation}

\noindent Note that the zero of energy is taken with vibrational zero-point energy taken into account. The electronic partition function taken to be the electronic spin multiplicity of the atom at the surface is:

\begin{equation}
    q_{el}=2S+1
\end{equation}

\noindent where $S$ is the total spin quantum number.

In our case the fine-structure energy levels discussed above are taken from high-accuracy experimental data \citep{NIST_ADS} and all levels below a wavenumber of 25000 cm$^{-1}$ above the electronic ground state are included. The typical area of a binding site is for all species assumed to be 0.2 nm$^2$, which seems like a reasonable estimate. The vibrational modes of the atom at the surface can be quite different depending on the type of binding. In our case a simple approximation has been used where all three frequencies are the same for each atom and it is also assumed that the force constant is the same for all atoms, meaning that relative magnitudes of frequencies between different atoms are proportional to the inverse of the square root of the mass ratio between the atoms. 

For Si typical vibrational frequencies are taken to have a wavenumber of 1000 cm$^{-1}$, and for example Fe vibrations will then have wavenumbers of 707 cm$^{-1}$ and C vibrations will have 1527 cm$^{-1}$. Upon desorption it is assumed that the remaining vibrational frequencies of the grain surface remain the same. Both this and the magnitude of the frequencies are relatively crude approximations that could be addressed with detailed analyses of vibrational frequencies. At this point, this introduces a level of complexity that is not really motivated considering other approximations in the models that have been employed. 

\begin{figure}
    \centering
    \includegraphics[width=\linewidth]{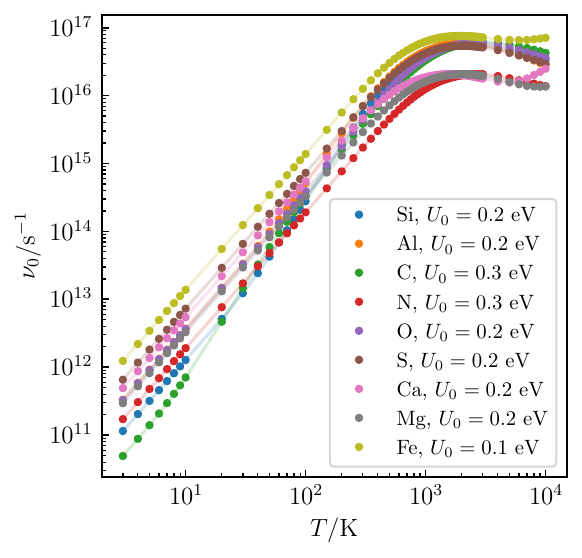}
    \caption{Desorption frequency factors and zero-point energies.}
    \label{fig:nu0}
\end{figure}

The multiplicities of the atoms at the surface have been taken to be the same as that of the gas-phase ground state. This means that for instance C, O, Si, and S atoms will have $2S+1=3$ (triplet state) and N atoms $2S+1=4$ (quartet state). It could be argued that those atoms probably form more strongly bound singlet and doublet states at the surface, respectively. This effort is also not motivated given the other uncertainties in the model. In the present manuscript, we did not take the spin polarization of the system into account. Obtaining a proper description of the electronic state of the reactant would require a detailed analysis of the electronic structure of the grain surface, which lies beyond the scope of this study. It should be mentioned that the actual rate of desorption, and thereby the frequency factor, will be affected by the rates of any transitions between electronic states that might be necessary as the atom leaves the grain surface. Including those effects would effectively lower the predicted desorption rate constants. Such detailed treatments are beyond the scope of the present study and it seems safe to conclude that the frequency factors calculated by the above procedure should be taken as an upper bound. These frequency factors are shown in Fig. \ref{fig:nu0}.   

\end{appendix}

\end{document}